\documentclass[prd,preprint,tightenlines,floatfix,showkeys,preprintnumbers,nofootinbib,eqsecnum]{revtex4}
 \usepackage[dvips,final]{graphicx}
  \usepackage{bm}
   \usepackage{epsf}
    \usepackage{epsfig}
     \usepackage{amssymb,amsthm,amsmath}
      \usepackage{latexsym}
       \usepackage{pifont}

\begin{document}

\title{Unitarization of elastic amplitude on $SO_{\mu}(2.1)$ group}

\author{O.N.Soldatenko$^a$}

\author{A.N.Vall$^a$}
 \email{vall@irk.ru}

\author{A.A.Vladimirov$^b$$^c$}
 \email{avlad@theor.jinr.ru}
  \affiliation{
  $~^a$Department of Theoretical Physics, Irkutsk State University, Irkutsk, 664003
  Russia \\ \\
  $^b$Bogoliubov Laboratory of Theoretical Physics,
  JINR, 141980, Moscow Region, Dubna, Russia\\ \\
  $^c$Institut f\"ur Theoretische Physik II,
Ruhr--Universit\"at Bochum, D--44780 Bochum, Germany}

\begin{abstract}
We obtain the solution of the unitarity equation for the elastic
processes in terms of the expansion coefficients of the amplitude
as a function on the $SO_{\mu}(2.1)$ group. This approach is a
generalization of the eikonal representation to the case of small
impact parameters and large transverse momenta. We show how the
unitarity relation is  modified when the contributions of the
backward scattering are taken into account. We discuss the
simplest models of the profile functions in the following cases:
full reflection, full absorption and the combination of these two
cases.
\end{abstract}


\maketitle

\section{Introduction}

Experimental data about elastic, inelastic and total
cross-sections of $pp$-scattering at energy $\sqrt{s}=14$ TeV will
be obtained at the LHC. The most interesting objects of study in
this energy region are the asymptotic behaviour of cross-sections
at high energy, approaching of the Froissart limit and the
behaviour of the ratio $\sigma_{el}/\sigma_{tot}$ at $s\rightarrow
\infty$. Also the mechanism of a possible deviation of that ratio
from the value $\sigma_{el}/\sigma_{tot}=1/2$ (black disk model)
is of interest.

The theoretical considerations to these problems are closely
related
 to the unitarization of the elastic scattering amplitude
\cite{Wall}, \cite{Troshin}. As a rule, for the unitarization of
an amplitude one uses the eikonal approximation. It is well known
that the eikonal approximation complies with the unitarity
equation for the elastic amplitude only in a region of small
scattering angles and large impact parameters.

In this paper, we obtain a solution of the unitarity equation in
terms of expansion coefficients of the elastic amplitude as a
function on the $SO_{\mu}(2.1)$ group, introduced in
\cite{Vall1}-\cite{artic2}. This approach allows us to generalize
the eikonal formalism of the impact parameter to the large angle
region and the small phase system volume of collided particles.

\section{Unitarity equation and its solution}

\begin{figure}[t]
\begin{center}
\includegraphics[scale=0.8] {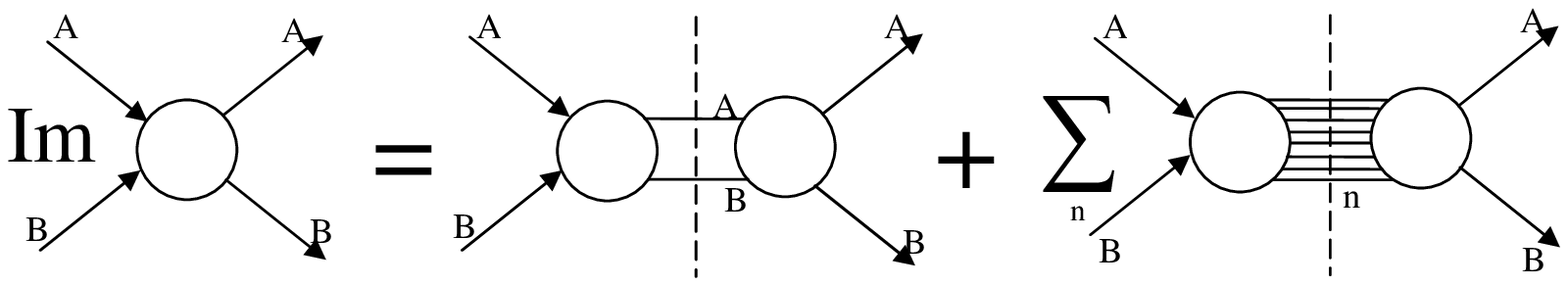}
\caption{\it The unitarity condition for the elastic process
$A+B\rightarrow A+B$}
 \label{unitarnost_1}
\end{center}
\end{figure}

At first, we note the main points of solution of the unitarity
equation in terms of the partial waves. Let us consider two
particle elastic process $A+B\rightarrow A+B$,
FIG.~\ref{unitarnost_1}. The unitarity equation is a consequence
of $S$ matrix unitarity -- $S^+S=1$:
\begin{equation}\label{est}
\mathrm{i}(F^{+}-F)=F^{+}F,~~~~\mathrm{i}F=S-I~.
\end{equation}
Using the expansion of the unity operator in Fock space and the
translation invariance of the operator $F$, as a result of
standard operations \cite{Collins} one obtains the equation on the
elastic amplitude $f^{\pm}(\vec{k}_{\bot};\vec{q})$:
\begin{equation}\label{vid_perepisi}
    2 Im f^{(+)}(\vec{q_{\bot}};\vec{p})=p^2\lambda (p)\sum \limits_{\epsilon=\pm 1}\int
    \mathrm{d}\Omega_{\vec{k}}\bar{f}^{(\epsilon)}(\vec{k_{\bot}};\vec{q})\
    f^{(\epsilon)}(\vec{k_{\bot}};\vec{p})\   +\
    A_{inel}(\vec{q},\vec{p})~,
\end{equation}
here
\begin{equation}\label{oboznach}
\begin{split}
&A_{inel}(\vec{q},\vec{p})= \sum \limits_{s}\int\prod \limits^{s}_{i=1}
     d\vec{k_i}\ \delta^4(P_{in}-\sum\limits_{i}k_i)\langle\{\vec{k}_i\}|A|\vec{q};-\vec{q}\rangle^*
     \ \langle\{\vec{k}_i\}|A|\vec{p};-\vec{p}\rangle, \\
 &f^{(\epsilon)}(\vec{k}_{\bot};\vec{q})\equiv
 \langle\vec{k}_{\bot},\epsilon \sqrt{q^2-k_{\bot}^2};
 \vec{k}_1=-\vec{k}|A|\vec{q};-\vec{q}\rangle,\\
 &\mathrm{d}\Omega_{\vec{q}} =
\frac{1}{q\sqrt{q^2-q^2_\bot }}\ \mathrm{d}\vec{q_\bot}~.
\end{split}
\end{equation}
$\epsilon$ is the sign of the projection of scattered particle
momenta on the z-axis, $\epsilon=\pm 1$. The first term on the
righthand side of equation (\ref{vid_perepisi})  describes a
two-particle contribution to the spectral density of the elastic
amplitude, FIG.~\ref{unitarnost_1}. The second term corresponds to
inelastic intermediate states. For derivation of
(\ref{vid_perepisi}) we use an integral relation which is valid
for an arbitrary function $\Phi(\vec{q})$:
\begin{equation}\label{Foc1}
  \int \mathrm{d}\vec{q}\ \ \Phi(\vec{q})= \int q^2\,\mathrm{d}q\, \mathrm{d}\Omega\, ~\Phi(\vec{q})=
  \sum\limits _{\epsilon=\pm 1}\int q^2 \mathrm{d}q \,
  \mathrm{d}\Omega_{\vec{q}}\ \ \Phi(\vec{q}_{\bot},\epsilon
  \sqrt{q^2-q_{\bot}^2})~.
\end{equation}

 Due to the momentum
conservation we have $p=|\vec{k}|=|\vec{q}|=|\vec{p}|$. The matrix
element of the $A$-operator is related to the matrix element of
the $F$-operator in the following way:
\begin{equation}\label{cvasAF}
  <f|F|in>=\delta^4(P_{in}-P_{f})<f|A|in>~.
\end{equation}
The first and the second arguments of
$f^{(\epsilon)}(\vec{k}_{\bot};\vec{q})$ correspond to the final
and the initial momenta of the scattered particle in the
center-of-mass system (c.m.s.) respectively. The factor
$\lambda(p)$ is related to the relative velocity in c.m.s. by a
relation:
\begin{equation}\label{skorost_scm}
|\vec{u}|=\frac{p(E_A+E_B)}{E_A\cdot E_B}=\frac{1}{\lambda (p)}~,
\end{equation}
where $E_A$, $E_B$ are the energies of colliding particles in
c.m.s.

Assuming $\vec{q}_{\bot}=0$ in the initial equation
(\ref{vid_perepisi}), it is easy to obtain the optical theorem:
\begin{equation}\label{optt}
 Im f^{(+)}(\vec{q}_{\bot}=0;\vec{p})= \frac{1}{8{\pi}^2
 \lambda(p)}(\sigma_{el}+\sigma_{inel})~,
\end{equation}
where
\begin{equation}\label{sigmanew}
  \sigma^{(\pm)}_{el}(p)=(2\pi)^2 p^2 \lambda^2(p)\int \mathrm{d}\Omega_{\vec{k}}
  \left|f^{(\pm)}(\vec{k}_{\bot}, \vec{p})\right|^2,
\end{equation}
\begin{equation}\label{sigmainel}
 \sigma_{inel}=(2\pi)^2 \lambda(p) \sum \limits_{s}\int\prod \limits^{s}_{i=1}
     d\vec{k_i}\
     \delta^4(P_{in}-\sum\limits_{i}k_i)|<\{\vec{k}_i\}|A|\vec{p};-\vec{p}>|^2~.
\end{equation}

Presently,  only one consistent method of solving equation
(\ref{vid_perepisi}) is known. This method is based on the
expansion of the amplitude
$f^{(\epsilon)}(\vec{k}_{\bot};\vec{q})$ over the representation
of the group $O(3)$, which is realized by Legendre polynomials
$P_l(z)$:
\begin{equation}\label{fOtri}
f(\vec{k};\vec{q})=\sum_{l=0}^{\infty}\ (2l+1)\ a_l(p)\
P_l\left(\frac{\vec{k}\cdot\vec{q}}{p^2}\right)~.
\end{equation}

Substituting expansion (\ref{fOtri}) into the unitarity equation
(\ref{vid_perepisi}), one obtains that $a_l(p)$ satisfies the
following
\begin{equation}\label{al}
  Im\ a_l(p)= 2 K\ |a_l(p)|^2 + g_{inel}(l,p),
\end{equation}
here
\begin{equation}\label{KOtri}
K=\pi\ p^2\lambda (p)~,
\end{equation}
\begin{equation}\label{etaOtri}
g_{inel}(l,p)=\frac{1}{4}\int \limits_{-1}^{1}\mathrm{d}c\ P_l(c)\
A_{inel}(\vec{q},\vec{p})~,~~c=\frac{\vec{q}\cdot\vec{p}}{p^2}~.
\end{equation}
Solving equation (\ref{al}) one obtains the following
representation for the partial amplitude $a_l(p)$:
$$
a_l(p)=\frac{\eta_l e^{2i\delta_l}-1}{4iK},\quad \text{with}\quad
g_{inel}(l,p)=\frac{1-\eta_l^2}{8K}~,
$$
where $\delta_l$ is an arbitrary real phase.

This is a well known result of the elastic amplitude unitarization
in terms of coefficients of its expansion on the $O(3)$ group.
Such description of scattering in terms of partial waves $a_l(p)$
is effective in low energy region, since the amplitude
$f(\vec{k};\vec{q})$ is saturated  by the lowest terms of
expansion (\ref{fOtri}). But, in a high energy region all terms in
the sum (\ref{fOtri}) become important, and the use of integration
instead of summation is needed. The procedure of changeover to the
integration corresponds to the eikonal quasi-classic approach:
$$
l\simeq bp, \qquad P_l\left(\frac{\vec{k}\cdot\vec{q}}{p^2}\right)\simeq J_0(bk_\perp)\ ,
 \qquad \mbox{at} ~\theta\sim 0, \quad bp\gg 1~,
$$
where $b$ has to be understood as an impact parameter, $k_\perp=p
\sin\theta$ and  $\cos\theta=\vec{k}\cdot\vec{q}/p^2$. Hence, the
expression for the amplitude (\ref{fOtri}) changes to the eikonal
representation:
\begin{equation}\label{econal}
   f(\vec{k};\vec{q})=2p^2\int\limits_0^\infty a(b,p)\ J_0\ (b k_\perp)\ b
   db~,
\end{equation}
here
\begin{equation}\label{abp}
    a(b,p)=\frac{\eta(b) e^{2i\delta(b)}-1}{4iK}~.
\end{equation}
In that case (\ref{econal}), $a(b,p)$  would satisfy condition
(\ref{al}).

However, here one has a problem: the eikonal amplitude
(\ref{econal}) with the profile function (\ref{abp}) does not
satisfy the initial unitarity equation (\ref{vid_perepisi}). The
reason for that is that Bessel functions $J_0\ (bk_\perp)$ do not
form an orthogonal system in the physical interval of values $0
\leq k_\perp \leq p\ $ i.e.
$$
\int\limits_0^p J_0\ (b_1k_\perp)\ J_0\ (b_2k_\perp)\ k_\perp
dk_\perp\ \neq\  \delta\ (b_1-b_2)~.
$$

In this paper, we attempt to solve the above mentioned problem in
the context of the theoretical-group generalization of the impact
parameter and obtain the unitarity equation in terms of profile
functions $u_p(b)$. In our formalism, the impact parameter $b$ is
interpreted as a vector of the maximal approach between two
particles in c.m.s.

Let us point the main steps of our formalism. First, we construct
a quantum analog of the impact parameter. As a basis we take the
expression for components of the vector of maximal approach
\cite{Vall1}:
\begin{equation}\label{8}
d_i=\frac{1}{q^2}\varepsilon_{ijk} q_j L_k~,
\end{equation}
where $q_j$ is a relative momentum of two particles, $L_k$ is a
relative orbital moment in c.m.s..

We take the expression (\ref{8}) as a basis for the construction
of a quantum analog of the impact parameter. The canonical
commutation relation between a relative momentum and a relative
coordinate lies in the basis of quantanization (here and later
$\hbar=1$):
$$[\xi _{i}\ q_{j}]=i\delta _{ij}~.$$
Then we perform the transition to the quantum description that
consists in the standard procedure of the substitution of the
$C$-numbers by the corresponding operators in the relation
(\ref{8}) and in the reduction of it to hermitian form. This leads
us to the following system of the commutation relations:
\begin{equation}\label{algebra1}
\begin{split}
&d_{i}=\frac{1}{q^{2}}(\varepsilon_{ijk}q_{j}L_{k}-iq_{i})~,\\
&d_{i}=(d_{i})^{+}\ \ ,\ \ [d_{i} q^2  ]=0, \quad [L_{i}d_{j}]=i\varepsilon_{ijk}d_{k}~,\\
&[L_{i}L_{j}]=i\varepsilon_{ijk}L_{k}, \quad
[d_{i}d_{j}]=\frac{-i}{q^{2}}\varepsilon_{ijk}L_{k}~.
 \end{split}
\end{equation}
The additial term in $d_i$ operator appears due to the requirement
of hermisity. The relations (\ref{algebra1}) show that operators
$d_{i}$ and $L_{j}$ form an algebra $SO(3.1)$ on the sphere
$q^2=const$. The Casimir operator of this algebra is a pure
number:
$$\hat{C}=d^2 -\frac{1}{q^2}L^2\equiv \frac{1}{q^2}~,$$
therefore we cannot use the full algebra (\ref{algebra1}) for the
construction of a wave function with defined impact parameter.

One way to build a wave function with defined impact parameter
consists in a selection of a nontrivial subalgebra of
(\ref{algebra1}) \cite{artic1}. The natural choice is
$d_{1},~d_{2},~L_{3}$ operators. They form the following algebra:
\begin{equation}\label{algebra2}
\begin{split}
[d_1 d_2]&=-\frac{i}{q^2} L_3 \\
[d_1 L_3]&=-i d_2 \\
[d_2 L_3]&=i d_1
 \end{split}
\end{equation}
This is the algebra of $SO(2.1)$ group. The properties of the
$SO(2.1)$ group are studied in detail in the monograph of
N.J.Vilenkin \cite{Vilenkin2}, and also in \cite{Vilenkin1}. This
group is noncompact and has the basic continued family of the
unitary representation and the discrete finite-dimensional family
of the nonunitary representation in the space of quadratically
integrable function. The Casimir operator of algebra
(\ref{algebra2}) is
\begin{equation} \label{algebra3}
\hat{K}=d^{2}_{\perp}-\frac{1}{q^2}L_3^2 \ \ ,\ \
d^{2}_{\perp}=d_1^2+d_2^2\ \ ,\ [d_{1,2}\hat{K}]=0\ ,\
[d_{3}\hat{K}]\neq 0~.
\end{equation}

As the next step, let us consider the following system of
equations
\begin{equation}\label{algebra18}
\hat{K}\Psi = b^2 \Psi \ \ ,\ \ L_3 \Psi = m \Psi~.
\end{equation}
In this equation we interpret the eigenvalue of the Casimir
operator $b^2$ as a squared impact parameter or as a squared
transverse component of the maximal approach vector of two
colliding particles.

Using the explicit form of operators $d_i~, L_3 $ in the momentum
space we obtain the particular solution at $m=0$:
\begin{equation}\label{algebra23}
  \Psi _{\mu} (\vec{q}_{\bot} )= \frac{q}{\sqrt{q^2 -q_{\bot}^2}}P_{-1/2+i\mu }\left ( \frac{q}{\sqrt{q^2 -q_{\bot}^2}} \right )
=\frac{1}{|\cos \theta|}P_{-1/2+i\mu }\left ( \frac{1}{|\cos \theta|} \right ),
\end{equation}
where $\theta$ is an azimuth angle of momentum $q$, $0 \leq \theta
\leq \pi $, $q_{\bot}=q \sin\theta$, and
\begin{equation}\label{muq}
\mu =\left ( q^2 b^2 -1/4  \right )^{1/2},~~ b^2 \geq
\frac{1}{4q^2}~.
\end{equation}
So, $\Psi _{\mu}(\vec{q}_{\bot})$ represents the wave function of
two-particle state in the momentum space with the defined value of
the squared impact parameter $b^2$. The spectrum restriction $q^2
b^2 \geq 1/4$ reflects the uncertainty relation for the phase
space of two-particle system. It limits small $b^2$ region at
fixed value of momentum $q$.

The connection between $\Psi _{\mu}(\vec{q}_{\bot})$ and the
eikonal representation kernel (\ref{fOtri}) follows from the Fock
expansion of the cone function \cite{Fok}:
\begin{equation}\label{algebra40++}
  P_{-1/2+i\mu }\left ( \frac{q}{\sqrt{q^2 -q_{\bot}^2}} \right )=J_0(bq_\bot)+O\left (\frac{q_\bot}{q}\right
  )+O\left (\frac{1}{bq}\right )~.
\end{equation}

As it was shown in \cite{artic1}, \cite{artic2}, the system of
functions $\Psi_{\mu}(\vec{q}_{\bot})$ forms an orthogonal and
complete basis independently for each of the two parts of the
momentum space $q_3>0$ and $q_3<0$ (so called forward and backward
semi-spheres):
\begin{equation}\label{ortog}
  \int \limits_0^{\pi /2}\Psi _{\mu} (\vec{q}_{\bot} )\Psi _{\nu} (\vec{q}_{\bot} )sin\theta d\theta =\frac{1}{\mu th(\pi
  \mu  )}\delta(\mu -\nu )~,
\end{equation}
\begin{equation}\label{polnota}
  \int \limits_0^{\infty}d\mu\ \mu th(\pi\mu)\Psi _{\mu} (\vec{q}_{\bot} )\Psi _{\mu} (\vec{q}_{\bot}')
  =\delta(|\cos \theta| -|\cos \theta'|)~.
\end{equation}

The analog of the Legendre expansion (\ref{fOtri}) is the
expansion of amplitude
$f^{(\epsilon)}(\vec{k}_{\bot};\vec{\kappa})$ as a function on the
group $SO_{\mu}(2.1)$ over the basis of
$\left\{\Psi_{\mu}(\vec{k}_{\bot})\right\}$:
\begin{equation}\label{amplituda_vvide}
  f^{(\epsilon)}(\vec{k_{\bot}};\vec{\kappa})= \int \limits_0^{\infty}(v_0 u^{'}_0)\
   P_{-1/2+i\mu}\ (v u^{'})\ u_p^{(\epsilon)}(\mu)\ \mu
   th(\pi\mu)d\mu~,
\end{equation}
where we introduce the cone variables:
\begin{equation}
 \begin{split}\label{konus}
v=(v_0,\vec{v})=\left(\frac{\kappa}{\kappa_3}, \frac{\vec{\kappa}_{\bot}}{\kappa_3}\right),&
\quad
v^2\equiv v_0^2-\vec{v}^{\ 2}=1, \quad \kappa_3>0\\
u^{'}=(u_0^{'},\vec{u}^{'})=\left(\frac{k}{|k_3|},\ \frac{\vec{k}_{\bot}}{|k_3|} \right),&
 \quad {u^{'}}^2 \equiv {u_0^{'}}^2-\vec{u}^{'2}=1, \quad |\kappa|=|k|=p\\
&(v u^{'})=v_0 u^{'}_0 - (\vec{v} \cdot\vec{u}^{'})~.
\end{split}
\end{equation}
We call  the expansion coefficient $u_p^{(\epsilon)}(\mu)$ in
(\ref{amplituda_vvide}) the profile function on the group
$SO_{\mu}(2.1)$.

Substituting the expansion (\ref{amplituda_vvide}) into the
initial equation (\ref{vid_perepisi}), we obtain the unitarity
condition for the profile functions $u_p^{(\epsilon)}(\mu)$ on the
group $SO_\mu(2.1)$:
\begin{equation}\label{localunitarn}
    Im\ u_p^{(+)}(\mu)=K \sum \limits_{\epsilon=\pm 1}
    |u_p^{(\epsilon)}(\mu)|^2+
    G_{inel}^{(+)}(\mu, p)~,
\end{equation}
where
\begin{equation}\label{Ginel}
 G_{inel}^{(+)}(\mu, p)=\frac{1}{4\pi}\int\
  \mathrm{d}\Omega_{\vec{q}}\Psi_{\mu}(q_{\bot})
  \ A_{inel}(\{\vec{q}_{\bot},
  q_3=+\sqrt{q^2-q^2_{\bot}}\},\vec{p})~.
\end{equation}

The equation (\ref{localunitarn}) is the main result of this
article. It differs from the unitarity relation for $a_l(p)$
(\ref{al}) by a presence of the signature $\epsilon$. The
appearance of signature $\epsilon$ allows us to consider
scattering amplitudes independently in forward and backward
semi-sphere.

In course of the derivation of (\ref{localunitarn}) we used the
summation theorem for spherical Legendre functions \cite
{Lebedev}, \cite{Gobson}:
\begin{equation}\label{tLegandra}
\begin{split}
  P_{-1/2+i\mu}&\{zz^{'}-(z^2-1)^{1/2}({z^{'}}^2-1)^{1/2}\cos\varphi\}=
  P_{-1/2+i\mu}(z)\ P_{-1/2+i\mu}(z^{'})+\\
  &+2\sum \limits_{m=1}^{\infty}(-1)^m\ \frac{\Gamma(-1/2+i\mu-m+1)}{\Gamma(-1/2+i\mu+m+1)}
  P_{-1/2+i\mu}^m (z)\ P_{-1/2+i\mu}^m (z^{'})\cos m\varphi~.
  \end{split}
\end{equation}

For convinience we rewrite $G_{inel}^{(+)}(\mu, p)$ from
(\ref{Ginel}) in the following parametric form:
\begin{equation}\label{pGinel}
 G_{inel}^{(+)}(\mu, p)=\frac{1-\eta_{inel}^2(\mu)}{4K}~.
\end{equation}
The parameter $\eta_{inel}(\mu)$ plays the role of a standard
inelastic coefficient and it varies within bounds  $ 0 \leq
\eta_{inel}(\mu) \leq 1$. The expression for cross-sections in
terms of the profile functions $u_p^{(\pm)}(\mu)$ follows from the
optical theorem (\ref{optt}) and the relation
(\ref{localunitarn}).
\begin{equation}\label{ssigma}
\begin{split}
&\sigma_{el}=\sigma_{el}^{(+)}+\sigma_{el}^{(-)} ~,\\
 &\sigma_{el}^{(\pm)}=(2\pi)^3 p^2 \lambda^2(p)
\int\limits_{0}^{\infty}  \left(|u_p^{(\pm)}(\mu)|^2\right)\mathrm{d}\Omega_{\mu}~,\\
&\sigma_{inel}=\frac{2\pi}{p^2}\int\limits_{0}^{\infty}
(1-\eta_{inel}^2(\mu))\mathrm{d}\Omega_{\mu}~.
\end{split}
\end{equation}

\section{Simple models}

As an application we discuss the simplest phenomenological models
for the profile function $u_p^{(\pm)}(\mu)$, which is consistent
with unitarity condition (\ref{localunitarn}).

Let us divide the interval of the impact parameter $b$ values in
three parts:
\begin{itemize}
  \item $0 \leq b \leq R_0(p)= 1/2p$ - the interval of forbidden $b$-values due to the condition $\ b^2 p^2 \geq 1/4$,
  \item $R_0(p)\leq b \leq R_{refl}(p)$ - the interval of full
  reflection,
  \item $R_{refl}(p)\leq b \leq R(p)$ - the interval of full
  absorption.
\end{itemize}

\begin{figure}[t]
\begin{center}
\includegraphics[scale=1.2] {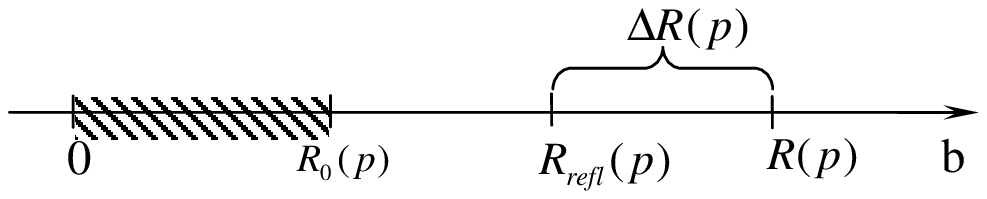}
\caption{\it The structure of the scattering area}
 \label{structura}
\end{center}
\end{figure}

\subsection{Full absorption}
First we assume that $R_{refl}(p)=R_0(p)$, which corresponds to
the full absorption target. In that case we have
 $$
 |u_p^{(-)}(\mu)|^2 \equiv 0\ , \quad \eta_{inel}=0~,
 $$
 in the whole area of the disk $R_0(p)\leq b \leq R(p)$. The
 unitarity condition (\ref{localunitarn}) takes the form
\begin{equation}\label{localunit}
    Im\ u_p^{(+)}(\mu)=K|u_p^{(+)}(\mu)|^2+ \frac{1}{4K}~.
\end{equation}
The solution of the equation  (\ref{localunit}) is
\begin{equation}\label{upl}
     u_p^{(+)}(\mu)=\frac{i}{2K}\ , \quad R_0(p)\leq b \leq R(p)~.
\end{equation}
Hence, for the cross-sections  $\sigma_{el}$ and $\sigma_{inel}$
we obtain:
\begin{equation}\label{sma}
\sigma_{el}=\sigma_{el}^{(+)} = \sigma_{inel}=\frac{2\pi}{p^2}
\int\limits_{0}^{\sqrt{R^2 p^2-1/4}} \mu th(\pi\mu)\
\mathrm{d}\mu~.
\end{equation}
For the case of $p^2R^2\gg1/4$ we find
\begin{equation}\label{stot}
\begin{split}
&\sigma_{el}=\sigma_{el}^{(+)} =\sigma_{inel}\simeq \pi R^2~,\\
&\sigma_{tot}=\sigma_{el}+\sigma_{inel}\simeq 2 \pi R^2~, \\
&\sigma_{el}/\sigma_{tot}=1/2~.
\end{split}
\end{equation}
Thus, using (\ref{localunitarn}) we obtain well known results for
the black disk model.

\subsection{Full reflection}
Now we consider the model of the full reflection disk, or in other
words let $R_{refl}(p)$ be equal $R(p)$. In this case we have
$$
G_{inel}=0\ , \quad R_0(p)\leq b \leq R(p)~.
$$
The unitarity relation (\ref{localunitarn}) takes the form
\begin{equation}\label{ravenstvo}
 Im\ u_p^{(+)}(\mu)= K
 \left(|u_p^{(+)}(\mu)|^2+|u_p^{(-)}(\mu)|^2\right)~,
\end{equation}
We rewrite this relation as follows
\begin{equation}\label{perepis}
 Im\ u_p^{(+)}(\mu)=K\
 |u_p^{(+)}(\mu)|^2+\frac{1-\eta_p^2(\mu)}{4K}~,
\end{equation}
where
\begin{equation}\label{eta}
  \eta_p(\mu)=\sqrt{1-|2K\ u_p^{(-)}(\mu)|^2}\ , \qquad  0\leq \eta_p(\mu) \leq
  1~.
\end{equation}

The solution of equation (\ref{perepis}) is
\begin{equation}\label{uplusmus}
 u_p^{(+)}(\mu)=\frac{\eta_p(\mu)\ e^{2i\delta(\mu, p)}-1}{2iK}~.
\end{equation}
where $\delta(\mu, p)$ is an arbitrary real phase. So, we see that
the amplitude of scattering into a backward semi-sphere
$u_p^{(-)}(\mu)$ plays a role of the effective absorption
coefficient for the scattering amplitude into forward semi-sphere
$u_p^{(+)}(\mu)$.

Let us define the absolute elastic  scattering on a disk with
radius $R$ as the scattering with parameters
\begin{equation}\label{uslovie}
\begin{split}
  &\eta_p(\mu)=0,\  \text{if}\ \   R_0(p)\leq b \leq R(p), \quad \text{(full reflection)} \\
  &\eta_p(\mu)=1,\  \delta(\mu, p)=0\ \  \text{if}\  b\geq R(p).
  \end{split}
\end{equation}
In this case we have:
\begin{equation}\label{sled_uslovie}
\begin{split}
  |&u_p^{(-)}(\mu)|^2=\frac{1}{4K^2},\quad u_p^{(+)}(\mu)=\frac{i}{2K},\quad \text{if}\ \   R_0(p)\leq b \leq R(p)~,\\
  &u_p^{(-)}(\mu)=u_p^{(+)}(\mu)=0,\quad \text{if}\ \   b \geq
  R(p)~.
\end{split}
\end{equation}
Then for the elastic scattering we obtain
\begin{equation}\label{sigmaintegral}
\sigma_{el}=\sigma^{(+)}_{el}+\sigma^{(-)}_{el}=\frac{4\pi}{p^2}\int
\limits_0^{\sqrt{R^2 p^2-1/4}}\mu\ th(\pi\mu)\ d\mu\  \simeq 2\pi
R^2~.
\end{equation}
And for the cross-section ratio we have
$$
\sigma_{el}/\sigma_{tot}\simeq 1~.
$$

\subsection{Combining model of reflection and absorption}
The integration interval for $\mu$ in cross-sections $\sigma_{el}$
and $\sigma_{inel}$ is divided into the interval of full
reflection ($0\leq \mu \leq \mu_{refl}$) and the interval of full
absorption ($\mu_{refl} \leq \mu \leq \mu_{R}$) as illustrated in
FIG.~\ref{structura}. Here
\begin{equation}\label{oblastmu}
\begin{split}
&\mu^2=0=p^2 R_0^2(p) -{1}/{4}~,\\
&\mu_{refl}^2=p^2 R^2_{refl}(p) -{1}/{4}~,\\
&\mu_{R}^2=p^2 R^2(p)-{1}/{4}~.
\end{split}
\end{equation}
Using obtained in the sections \textbf{III.A} and \textbf{III.B}
 expressions for $u_p^{(+)}(\mu)$, $u_p^{(-)}(\mu)$,
$\eta_{inel}$  and the expression for cross-section
(\ref{ssigma}),  it is easy to find that
\begin{equation}\label{chastn}
\sigma_{el}^{(+)}=\frac{2\pi}{p^2}\int\limits_{0}^{\mu_R}
\mathrm{d}\Omega_{\mu}\ ,\quad
\sigma_{el}^{(-)}=\frac{2\pi}{p^2}\int\limits_{0}^{\mu_{refl}}
\mathrm{d}\Omega_{\mu}\ ,\quad
\sigma_{inel}=\frac{2\pi}{p^2}\int\limits_{\mu_{refl}}^{\mu_R}
\mathrm{d}\Omega_{\mu}~.
\end{equation}
Thus, for the cross-section ratio we obtain
\begin{equation}\label{chastotn}
 \sigma_{el}/\sigma_{tot}=1-\Delta~,
\end{equation}
where
\begin{equation}\label{Delta}
\Delta=\frac{1}{2}\ \frac{ \int\limits_{\mu_{refl}}^{\mu_R}
\mathrm{d}\Omega_{\mu}}{ \int\limits_{0}^{\mu_R}
\mathrm{d}\Omega_{\mu}}~.
\end{equation}
The values of $\Delta$ lie within the range  $0\leq \Delta \leq
1/2$. The limit value $\Delta=0$ is reached for $
R_{refl}\rightarrow R$, which corresponds to the case of the full
reflection. The limit value $\Delta=1/2$ is reached for $
R_{refl}\rightarrow R_0$, which corresponds to the case of the
full absorption. In the region of the phase valume $R^2 p^2 ,
R_{refl}^2\ p^2 \gg 1/4$, we can obtain:
\begin{equation}\label{chaotn}
 \sigma_{el}/\sigma_{tot} \simeq \frac{1}{2}\ (\ 1+ \frac{R_{refl}^2}{ R^2})\ ,\  R_{refl}\leq
 R~.
\end{equation}

\begin{figure}[t]
\begin{center}
\includegraphics[scale=0.6] {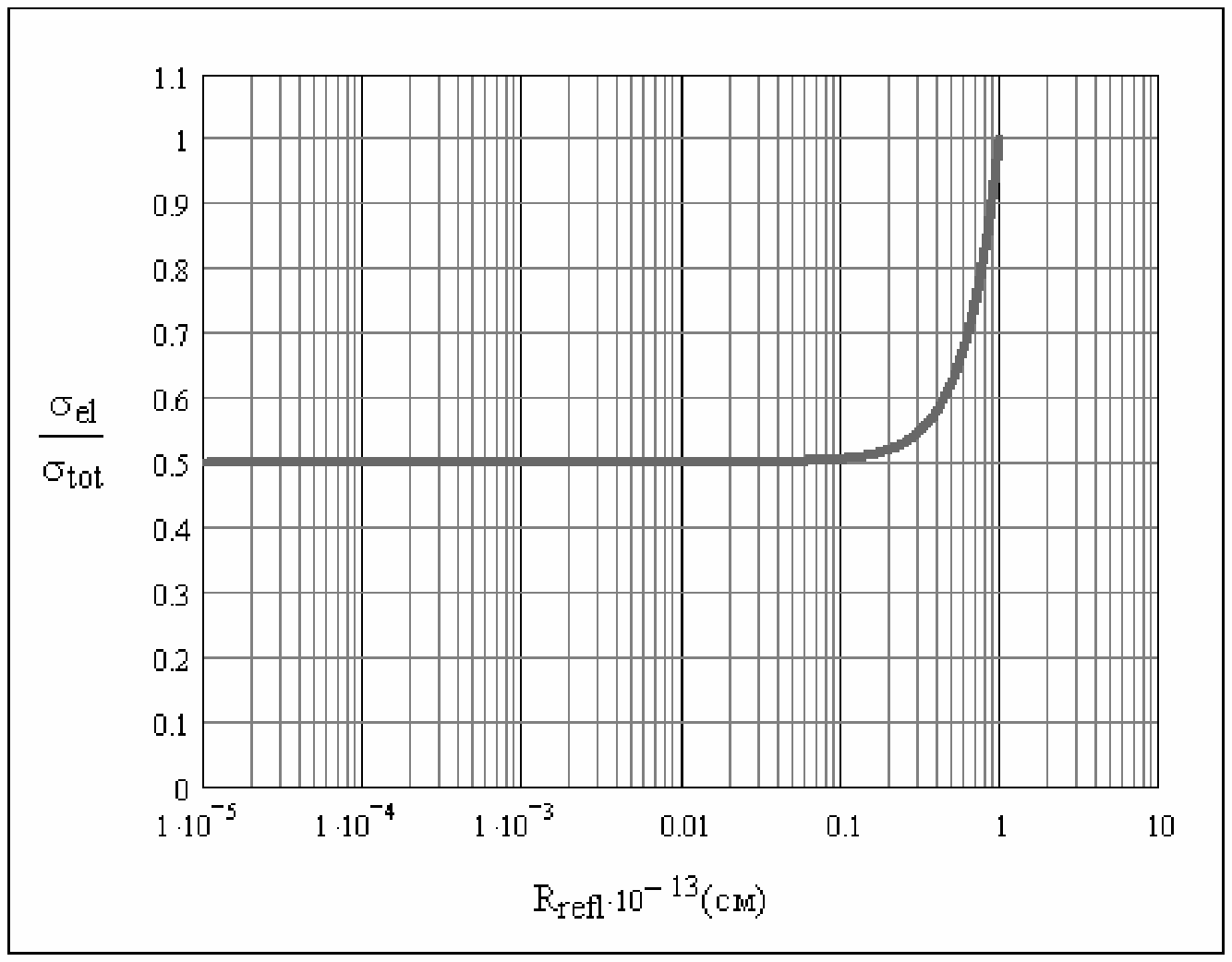}
\caption{\it The dependence of the ratio
$\sigma_{el}/\sigma_{tot}$ from $R_{refl}$}
 \label{otnoshenieS}
\end{center}
\end{figure}

The results of the numerical calculations of integrals
(\ref{chastotn}) at $\sqrt{s}=14$ TeV and $R=1$F are shown in
FIG.~\ref{otnoshenieS}. One can see that the deviation of the
ratio $\sigma_{el}/\sigma_{tot}$ from the value of $1/2$ starts in
the region, where $R_{refl}$ is of order $R$. In the region of
$R_{refl}\sim R$ the value of $\sigma_{el}/\sigma_{tot}$ changes
sharply and converges to 1. Thus, the unitarity saturation takes
place on an object with a very narrow full absorption periphery,
$\Delta R(p)/R \ll 1$. In other words, the unitarity saturation
results from the processes of scattering to backward semi-sphere
(also called Reflective Scattering).

There is a prediction for the ratio of cross-section
$\sigma_{el}/\sigma_{tot}$ reaching the value of $0.67$ at
 LHC energies $\sqrt{s}=14$ TeV \cite{TTModel}. For these values, from
relation (\ref{chastotn}) we have  $R_{refl} \simeq 0.58$F. Thus,
we conclude that, for the energies $\sqrt{s}=14$ TeV, the size of
the full reflection region is almost of the same order as the size
of the full absorption region.

\section{Conclusion}
In the present paper, we have shown that the unitarity equation
for the elastic processes can be rewritten in terms of the profile
functions $u^{\pm}_p(b)$, where $b$ is the generalization of the
impact parameter in the context of the group-theoretical approach
\cite{artic1}. In such terms the unitarity equation is local for
the whole interval of $b$, $1/2p\leq b \leq \infty$, where $p$ is
the momentum in c.m.s. The signature $\epsilon=\pm$ of the
scattering into the forward and the backward semi-sphere appears
in the unitarity equation. The appearance of signature $\epsilon$
allows us to consider scattering amplitudes independently in
forward and backward semi-sphere.

The analysis of the dependence of the ratio
$\sigma_{el}/\sigma_{tot}$ on the radius of the full reflection
$R_{refl}(p)$ and the diffraction radius $R(p)$ is given within
the limits of the simplest models for the profile function
$u_p(\mu)$. It is shown that the limit of the unitarity saturation
$\sigma_{el}/\sigma_{tot}\rightarrow 1$ is achieved at the limit
case of full reflection, $R_{refl}(p)\rightarrow R(p)$.

\section{Acknowledgments}
We are grateful to prof. S.E. Korenblit for the discussions,
critical remarks and useful suggestions.

The work was supported by the grant the president of Russian
Federation for the support of the leading scientific schools (NSh
-1027.2008.2.)

\end{document}